\documentclass[11pt]{article}
\usepackage{amssymb}
\usepackage{amscd}
\usepackage[dvips]{graphicx}
\usepackage{yfonts}
\usepackage{txfonts}
\usepackage{wasysym}
\begin{document}
\begin{flushright}
       University in Bia{\l}ystok, July 2019\\
\end{flushright}
\bigskip
\bigskip
\begin{center}
\Large{The spinning particles -  classical description}
\end{center}
\bigskip
\bigskip
\begin{center}
{\large{\textbf{Cezary J. Walczyk}} }\\
Department of Physics\\
University in Bia{\l}ystok,
ul. Ciolkowskiego 1L, 15-245 Bia{\l}ystok, Poland\\
\end{center}
\begin{center}
\large{  \textbf{Zbigniew Hasiewicz}}\\
{Department of Physics\\ University in Bia{\l}ystok,
ul. Ciolkowskiego 1L, 15-245 Bia{\l}ystok, Poland\\ }
\end{center}
\bigskip\bigskip
\date{July 2019}
\begin{abstract}
\noindent The classical model of spinning particle is analyzed in details in two versions - with single spinor and two spinors put on the trajectory.  Equations of motion  of the first version are easily solvable. The system with two spinors becomes non-linear. Nevertheless the equations of motion are analyzed in details and solved numerically. In either case the trajectories are ilustrated and their properties are disussed. There is also discussion of possible physical quantities associated with the spinning motions. Among others: the size of particles and their gyromagnetic ratios. 
Finally, some possible, speculative explanations of the properties of the Universe are proposed: the origin and nature  of dark matter and lack of the equilibrium bettween mater and anti-matter.  
\end{abstract}
\vspace{1.5cm}
\thispagestyle{empty}
\eject
\section*{Introduction}

The  paper is devoted to the presentation and detailed analysis  of the classical model of point particle carrying spinor variables. The classical models of spinning particles  were introduced a long time ago (\cite{CII:6,Rivas_1,Rivas_2,barut1,CII:7}). Despite of their apparent simplicity they were never analyzed in full extent. The present paper is devoted to fill this gap. In the course of this task the authors learned that the classical motions of spinning particles are interesting and promising enough to present them in the scientific paper. First of all it is possible 
to make strict connection of strictly mathematical entity of spinor and physical notion of spin. In fact the classical spin variable is used to solve the linear and non-linear equations of motion for spinning particles.  
The action functionals governing their motion via Lagrange equations, is a natural generalization of point particle action by putting spinor variables on the trajectories. The spinors are "minimally coupled" to the trajectory. The action is supplemented by kinetic term for spinors. It is of first degree in their time derivatives. The system is obviously a constrained one but due to the last property, the constraints are of mixed class. This feature plays crucial role in the description of the models on the quantum level. The appropriate modification of BRST approach results \cite{anom_coho1, anom_coho2} in description of classical fields with arbitrarily high spin together with the strict recipe for construction of their Lagrange functions. Despite of this remark the quantum counterpart of the models will not be considered here. 
The discussion of quantum properties will be postponed to the future publication by the authors. 
The systems are (by construction) Poincare invariant. This enables to perform the analysis and  calculations in the  conventional rest frame. 
It should be stressed that the spinor variables considered here are real i.e. of Majorana type.\\
The paper is organized as follows.  In the first section the (linear) system with single spinor put on the trajectory is introduced and analyzed in details. In contrast to conventional wisdom the  physical spin is consistently identified already on classical level. It appears that the spin variables are of great value in description of the classical motion.  The magnetic moment and dipole moment are identified too. There is also a comment on the impossibility to determine the proper gyromagnetic ratio within the classical picture. Finally   the trembling motion (Zitterbewegung) is discussed briefly.  In the second chapter the nonlinear system containing two spinor variables is anlyzed in details. The nonlinearity is the price one has to pay for CPT invariance. The complex system of two spinors interacting 
via common trajectory is described in terms of individual spin ingredients - each corresponding to different spinor.
The parameterization of the trajectories of two spinor  particle is changed and adopted to a non-linear problem. This movement allows one to find all solutions of non -linear differential equations  in analytic way. The analysis of closed trajectories (in the rest frame) is presented and some of their graphs are presented. There is also discussion of the mass spectrum of the system - it appears  fuzzy continuous even in the case of luxon particle. This phenomenon is completely different from the case of single spinor particle where the mass trajectories connect  discreet sets of points corresponding to integral or half-integral spins.
 \eject

\section{ The single spinor system }

The spinning particle model is defined by the action functional being the natural modification of free relativistic particle whose action functional 
\begin{equation}
\mathcal{A}_0=-m_0\int d\tau\sqrt{-g_{\nu\mu} \frac{d x^\nu}{d\tau} \frac{d x^\mu}{d\tau}}.\label{sample}
\end{equation}
being prportional to the  proper time of the point object of mass $m_0$. The above functional is reparametrization invariant i.e. under smooth change of the trajectory parameter
\[
\tau\rightarrow\tau '=f(\tau).
\]
One may use equivalently the action which is quadratic in time derivatives, at the price of introducing
subsidiary $1$-bein  variable $e$ (Lagrange multiplier) which has the nature of nowhere vanishing $1$-form along the trajectory and transforms under reparametrization as follows
$\;
e\rightarrow e'=e\left(\frac{df}{d\tau}\right)^{-1}.
\;$
Then the action takes the form
\begin{equation}
\mathcal{A}_0=\frac{1}{2}\int d\tau \left(e^{-1}g_{\mu\nu}\dot{x}^\mu\dot{x}^\nu-e m_0^2\right).
\label{action e}
\end{equation}

The natural extension of the above functional to include the spinor variables put on the trajectory is
defined to be 
\begin{equation}
\mathcal{A}_I=\int d\tau \mathcal{L}_I=\mathcal{A}_0+\frac{1}{2}\int d\tau\left[(-1)^{I+1} h\dot{x}j_I+2g \bar{\eta}_I\dot{\eta}_I\right], \quad I\in \{1,2\},
\label{lagrange spinor}
\end{equation}
where
\begin{equation}
\label{Spinor bilinears}
j_I^\mu=\eta_I^\alpha[C\gamma^\mu]_{\alpha\beta}\eta_I^\beta,\quad\quad
\bar{\eta}_I\dot{\eta}_I=\eta_I^\alpha C_{\alpha\beta}\dot{\eta}_I^\beta. 
\end{equation}
are bilinear functions in spinors. The index $I$ allows one to consider two essentially different (under assumption $h\geq 0)$ couplings of the spacetime trajectory with spinors. In the first section the variable $I$ is fixed to take one of the possible values. The coupling parameters $g$ and $h$ are introduced to make it easier to watch the influence of the particular terms in (\ref{lagrange spinor})
on the motion.
It should be stressed that the Majorana spinors do exist for the flat spacetime metric 
$g = { \rm diag}(-1,1,1,1) $ and the matrix $C$, defining $S\!O(3,1)$ invariant form  present in 
(\ref{Spinor bilinears}) is absolutely  antisymmetric.\\
Passing to the analysis of the equations of motion generated by (\ref{lagrange spinor}) it is worth to determine the canonical momenta conjugated to the variables  $x^\mu$, $\eta_I^\alpha$ and $e$:
\begin{eqnarray}
p_{I\mu}&=&\frac{\partial \mathcal{L}_I}{\partial \dot{x}^\mu}=e^{-1}\dot{x}_\mu-\frac{h}{2}(-1)^I \mathrm{j}_{I\mu},\label{momentum x}\\
\pi_{I\alpha}&=&\frac{\partial \mathcal{L}_I}{\partial \dot{\eta}^{\alpha}}=-g\eta_{I\alpha}\label{momentum spin},\\
\pi_{e}&=&\frac{\partial\mathcal{L}_I}{\partial \dot{e}}=0,\label{momentum e}
\end{eqnarray}
It is clear that the system defined by the action (\ref{lagrange spinor}) is  a constrained one. 
Nevertheless the Lagrange equations can be solved:
\begin{eqnarray}
0=\dot{\pi}_{e}&=&-\frac{1}{2}\left(e^{-2}\dot{x}^2+m_0^2\right),\label{dot e}\\
\frac{d}{d\tau}\left(e^{-1}\dot{x}_\mu-\frac{h}{2}(-1)^I \mathrm{j}_{I\mu}\right)=\dot{p}_{\mu}&=&0,\label{dot pi}\\
-g\dot{\eta}_{I\alpha}=\dot{\pi}_{I\alpha}&=&g\dot{\eta}_{I\alpha}-h(-1)^I \dot{x}_\nu [C\gamma^\nu]_{\alpha\beta}\eta_{I}^\beta\label{dot eta}.
\end{eqnarray} 
The choice of standard gauge $e=1$ reduces (\ref{dot e}) to the standard relation:
\begin{equation}
\dot{x}^2=-m_0^2,\label{mass relation}\;.
\end{equation}
From the equation (\ref{dot pi})it is evident that the canonical momenta are constants of motion. Fixing their values one may express $\dot{x}^\mu$ in terms of  $\eta_{I}^\alpha$ variables. Taking into account this relation in (\ref{dot eta}) allows one to write down the system of equations for spinor variables:
\begin{equation}
g\dot{\eta}_{I\alpha}= \frac{h}{2}(-1)^I p_{\nu}[C\gamma^\nu]_{\alpha\beta}\eta_{I}^\beta+\frac{h^2}{4} \mathrm{j}_{I\nu}[C\gamma^\nu]_{\alpha\beta}\eta_{I}^\beta.
\label{spinor equation} 
\end{equation}
It is rather well known that the vector $\mathrm {j}_{I}$ (\ref{Spinor bilinears}) constructed out of single Majorana spinor is lightlike $\mathrm{j}_{I}^2 = 0$.  For Majorana spinors even stronger identity is satisfied:
$$
\mathrm{j}_{I\nu}{[C \gamma^\nu]}_{\alpha\beta}\eta_{I}^\beta=0.
$$
For this reason the system of equations (\ref{spinor equation}) simplifies to linear ones
\begin{equation}
g\dot{\eta}_{I}^{\alpha}= \frac{h}{2}(-1)^I p_{\nu}{[\gamma^\nu]^\alpha}_\beta\eta_{I}^\beta,\label{spinor linear} 
\end{equation}
with solution given by exponent:
\begin{equation}
{\eta_I^\alpha}(\tau)={{[e^{\frac{h}{2g}(-1)^I p_\nu \gamma^\nu\tau}]}^\alpha}_\beta \eta_I^\beta(0).\label{spinor linear solution}
\end{equation}
Using the Clifford relation of gamma matrices one may transform the exponent to trigonometric form
\begin{equation}
e^{\frac{h}{2g}(-1)^I p_\nu \gamma^\nu \tau}=\cos\frac{ h m}{2g}\tau +(-1)^I\frac{p_\nu\gamma^\nu}{m}\sin\frac{h m}{2g}\tau,
\end{equation}
which gives finally:
\begin{equation}
\eta_I^\alpha(\tau)=\eta_I^\alpha(0)\cos\frac{ h m}{2g}\tau +(-1)^I\frac{1}{m}p_\nu{[\gamma^\nu]^\alpha}_\beta\eta_I^\beta(0)\sin\frac{ h m}{2g}\tau .
\label{eta solution}
\end{equation}
Here and in the sequel $m = \sqrt{-p_\mu p^\mu}$. This function is real for massive momenta, purely imaginary for tachyonic ones and zero for massless particle. In the case of tachyons the trigonometric 
functions above  are changed into hyperbolic ones. For the lightlike momenta the evolution is linear in $\tau$.\\ 
In the sequel only massive case $m^2 > 0$ will be under consideration.\\
Using (\ref{eta solution}) and the standard definition of basis bivectors  $\gamma^{\mu\nu}=[\gamma^\mu,\gamma^\nu]/2$ one finds an evolving vector build out of the spinors:
\begin{eqnarray}
\mathrm{j}_I^\mu(\tau)&=-\frac{p^\mu p_\nu}{m^2}\mathrm{j}_I^\nu(0)+\left(\delta^\mu_\nu+\frac{p^\mu p_\nu}{m^2(p)}\right)\mathrm{j}_I^\nu(0)\cos\frac{h m}{g}\tau\nonumber\\
&+(-1)^I\eta_I^\alpha(0)[\gamma^{\mu\nu}]_{\alpha\beta}\eta_I^\beta(0)\frac{p_\nu}{m}\sin\frac{h m}{g}\tau .\label{evolving current}
\end{eqnarray}
Taking into account the above and the equation (\ref{momentum x}) one finds finally 
\begin{eqnarray}
x_I^\mu(\tau)&=\frac{1}{2}p^\mu\left(1+\frac{m_0^2}{m^2}\right)\tau+\frac{g}{2m}(-1)^I\left(\delta^\mu_\nu+\frac{p^\mu p_\nu}{m^2}\right)\mathrm{j}_I^\nu(0)\sin\frac{h m}{g}\tau\nonumber\\
&-\frac{g}{2m}\eta_I^\alpha(0)[\gamma^{\mu\nu}]_{\alpha\beta}\eta_I^\beta(0)\frac{p_\nu}{m}\cos\frac{h m}{g}\tau+x_{I}^\mu(0),\label{x solution}
\end{eqnarray}
where the integration constants $x_{I}^\mu(0)$ are determined by initial conditions. The solution (\ref{x solution}) is a superposition of a periodic motion generated by spinors and translational motion in the direction of the conserved momentum $p^\mu$. In order to obtain clear physical and geometrical interpretation of the motion it is convenient to perform further analysis in the inertial reference frame where $(p^0,\vec{p})=(m,\vec{0})$. This frame will be conventionally called the "rest frame". Introducing a notation $(\mathrm{x}_I^{0},\vec{\mathrm{x}}_I)$, $(\mathrm{j}_I^{0},\vec{\mathrm{j}}_I)$ for the respective variables referred to this frame  and defining a vector  $\vec{\mathrm{k}}_I$ and frequency $\Omega_p$
\begin{eqnarray} 
\vec{\mathrm{k}}_I=\frac{1}{2}\eta_I^{(0)\alpha}(0)[\vec{\gamma},\gamma^{0}]_{\alpha\beta}\eta_I^{(0)\beta}(0),
\;\;\;\Omega_p=\frac{hm(p)}{g},\label{rest frame quantities}
\end{eqnarray}
one obtains the evolution in chosen rest frame:
\begin{eqnarray}
\mathrm{j}_I^{0}(\tau)&=&\mathrm{j}_I^{0}(0)=\mathrm{const.} ,\label{zero spinor curr}\\
\vec{\mathrm{j}}_I(\tau)&=&\vec{\mathrm{j}}_I(0)\cos\Omega_p\tau-(-1)^I\vec{\mathrm{k}}_I\sin\Omega_p\tau,\label{vect spin curr}\\
\dot{\mathrm{x}}_I^{0}(\tau)&=&m(p)+\frac{h}{2}(-1)^I \mathrm{j}_I^{0}(0),\label{zero x der }\\
\dot{\vec{\mathrm{x}}}_I(\tau)&=&\frac{h}{2}\left((-1)^I \vec{\mathrm{j}}_I(0)\cos\Omega_p\tau-\vec{\mathrm{k}}_I\sin\Omega_p\tau\right), \label{vest x der}\\
\mathrm{x}_I^{0}(\tau)&=&\left(m(p)+\frac{h}{2}(-1)^I \mathrm{j}_I^{0}(0)\right)\tau, \label{x zero ev}\\
\vec{\mathrm{x}}_I(\tau)&=&\frac{g}{2m(p)}\left((-1)^I \vec{\mathrm{j}}_I(0)\sin\Omega_p\tau +\vec{\mathrm{k}}_I\cos\Omega_p\tau\right)+ \vec{\mathrm{x}}_{Ic}.\label{vec x ev}
\end{eqnarray} 
From the above it is clearly seen that the equations define the motion in the plane determined by the vectors $\vec{\mathrm{j}}_I(0)$ and $\vec{\mathrm{k}}_I$. Taking into account the identity $\mathrm{j}_{I\nu}(\tau)\mathrm{j}_I^\nu(\tau)=0$ one may conclude that $|\vec{\mathrm{j}}_I(0)|=|\vec{\mathrm{k}}_I|=\mathrm{j}_I^{0}(0)$ and $\vec{\mathrm{j}}_I(0)\perp\vec{\mathrm{k}}_I$.This in turn means that the particle moves in the rest frame along the circles of the radius $\mathfrak{r} _I$ given by
\begin{equation}
\mathfrak{r}_I=\frac{g}{2m(p)}|\mathrm{j}_I^{0}(0)|,\label{radius}
\end{equation}
with linear speed
\begin{equation}
\mathrm{v}_I=\frac{|\dot{\vec{\mathrm{x}}}_I(\tau)|}{ \dot{\mathrm{x}}_I^{0}(\tau)}=\frac{h}{2}\frac{|\mathrm{j}_I^{0}(0)|}{m(p)+\frac{h}{2}(-1)^I \mathrm{j}_I^{0}(0)}.\label{speed}
\end{equation}
The centre of the circle $\vec{\mathrm{x}}_{I}(0)$ is determined by initial conditions.\\
The equations (\ref{zero spinor curr}), (\ref{momentum x}) and (\ref{mass relation}) allow one to determine the dependence of the mass $m$ on the constant $\mathrm{S}_I=\frac{g}{2}\mathrm{j}_I^{0}(0)$:
\begin{equation}
m = \sqrt{\frac{h^2\mathrm{S}_I^2}{g^2}+{m_0}^2}-\left(-1\right)^I \frac{h\mathrm{S}_I}{g}.
\label{mass spin}
\end{equation}
In the next subsection it will become clear that the quantity $\mathrm{S_I}$ is the length of the vector being the classical equivalent of spin. Nevertheless this quantity will be called spin from now on.
\begin{figure}[ht]
\begin{center}
\includegraphics[width=8.cm]{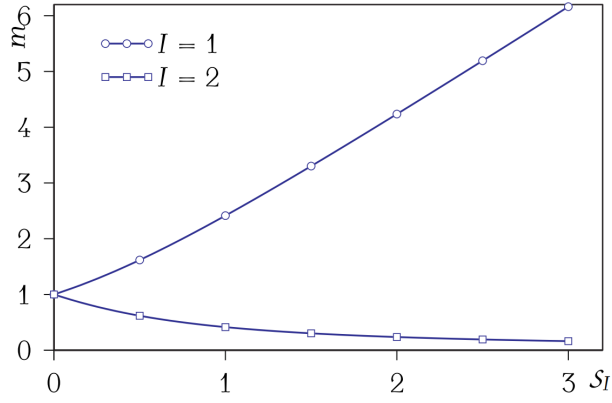}
\caption{Mass - spin $\mathrm{S}_I , I = 1,2$ dependence. Figures were drawn for $m_0 = h = g =1$.}
\label{fig.1.1}
\end{center}
\end{figure}
Depending on $I = 1,2$ the mass is growing and decreasing  function of spin respectively (fig.\ref{fig.1.1}).

\begin{figure}[ht]
\begin{center}
\includegraphics[width=13.cm]{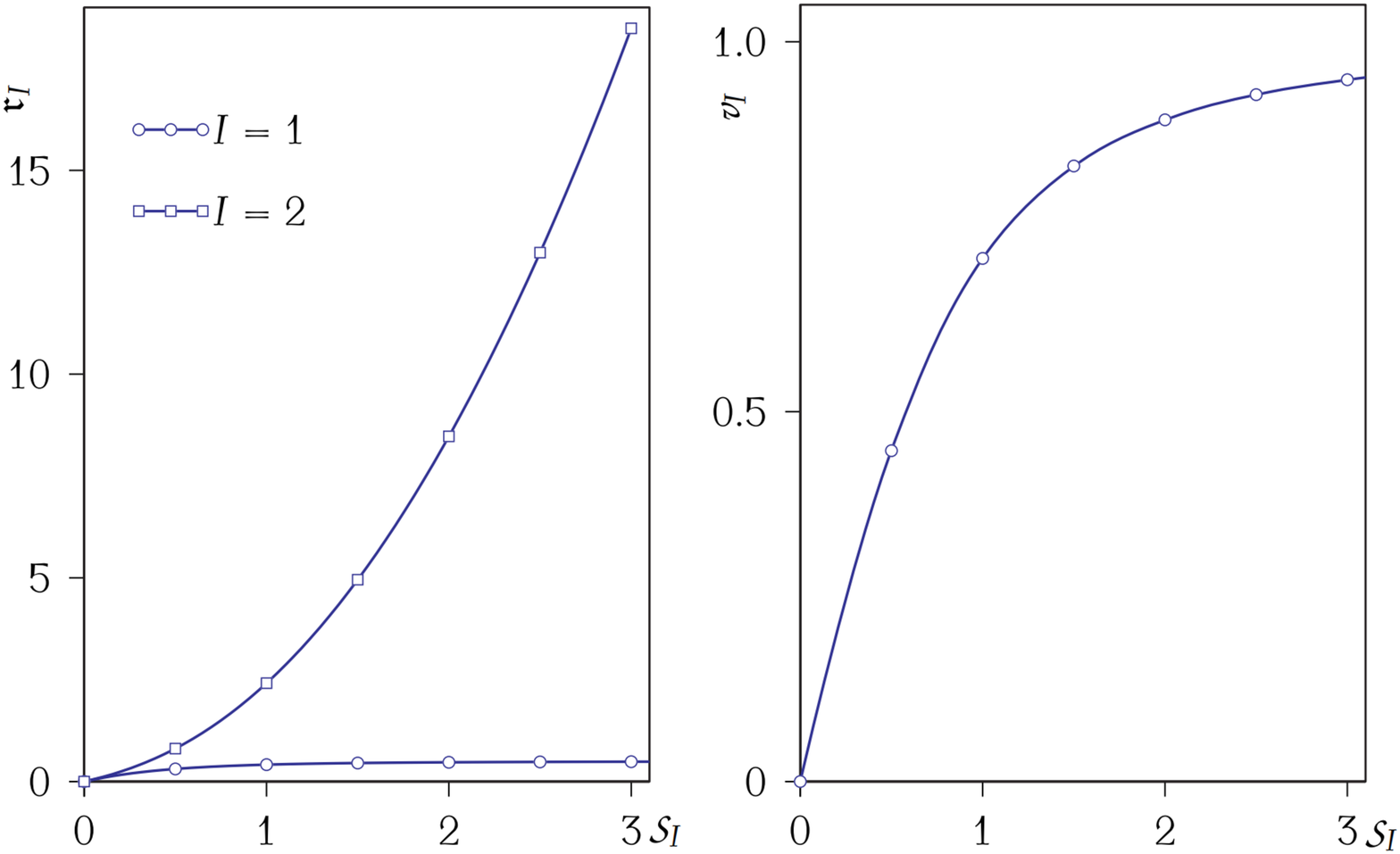}
\caption{The dependence  of the radius $\mathfrak{r}$ and speed $\mathrm{v}$ on spin $\mathrm{S}_I$. Figures are drawn for $ m_0 = h =g =1$.}
\label{fig.1.2}
\end{center}
\end{figure}

\noindent The variant dependence of the masses on spin for $I = 1,2$ has its reflection in the sizes of the particle orbits (fig.\ref{fig.1.2})
\begin{equation}
\mathfrak{r}_I=\frac{\mathrm{S}_I}{m(p)},
\label{size}
\end{equation}
whereas the speed is independent of the value of $I$ i.e. on the sign of the coupling of spinors with trajectory 
\begin{equation}
\mathrm{v}_I=\frac{1}{g}\frac{\mathrm{S}_I}{\sqrt{\left(\frac{\mathrm{S}_I}{g}\right)^2+\left(\frac{m_0}{h}\right)^2}}.
\label{speed ind}
\end{equation}

\subsection{Spin}
\begin{figure}[ht]
\begin{center}
\includegraphics[width=9.cm]{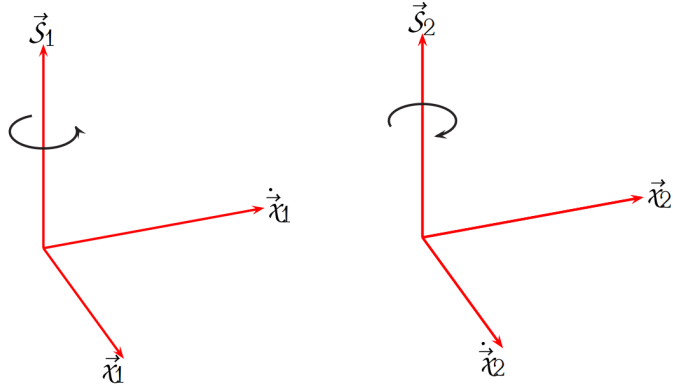}
\caption{Orthogonal systems where $I \in\{1,2\}$.}
\label{fig.1.3}
\end{center}
\end{figure}

From Noether theorem one obtains the conserved quantities (components of angular momentum) corresponding to invariance of the action under Lorentz transformations:
\begin{eqnarray}
\mathcal{J}_I^{\mu\nu}=p^\nu x^\mu-p^\mu x^\nu \underbrace{+\frac{1}{2}\pi_I^\alpha[\gamma^{\mu\nu}]_{\alpha\beta}\eta_I^\beta}_{Z_I^{\mu\nu}}, \;\;\;\; Z_I^{\mu\nu}Z_{I\mu\nu}=0.\label{ang momentum}
\end{eqnarray}
The last identity is satisfied provided (\ref{momentum spin}) is taken into account.
The purely spatial part of the above tensor defines a pseudovector 
\begin{equation}
J_I^i=\frac{1}{2}\varepsilon_{ijk} \mathcal{J}_I^{jk},\quad \mathrm{where}\quad i,j,k\in \{1,2,3\}.
\label{angular momentum spatial}
\end{equation}
which is interpreted as angular momentum. In the course of analysis of the spinorial system the point 
pointed by the vector $\vec{x}_c$ was distinguished and called the center of mass. It is then natural to introduce the coordinates with respect to this point. The tensor (\ref{ang momentum}) is decomposed accordingly:
\begin{eqnarray}
\mathcal{J}_I^{\mu\nu}=\underbrace{p^\nu x_c^\mu-p^\mu x_c^\nu}_{L_I^{\mu\nu}}+ \underbrace{p^\nu (x^\mu-x_c^\mu)-p^\mu (x^\nu-x_c^\nu)+Z_I^{\mu\nu}}_{S_I^{\mu\nu}},\label{ang momentum dec}
\end{eqnarray}
In order to make the definition complete it is worth to mention that  $x_c^0$ is a time of the mass centre. Applying the rule above one may split (\ref{angular momentum spatial}) into a sum of angular momentum of that mass centre $\vec{L}_I$ and spin momentum $\vec{S}_I$:
\begin{eqnarray}
J_I^i=\underbrace{\frac{1}{2}\varepsilon_{ijk} L_I^{jk}}_{L_{I}^i}+ \underbrace{\frac{1}{2}\varepsilon_{ijk} S_I^{jk}}_{S_{I}^i},\quad \mathrm{where}\quad i,j,k\in \{1,2,3\}.\label{angular momentum spatial split}
\end{eqnarray}
It is worth to compare the the quantity called "spin" with invariants of Poincare group. There are two independent of them. The first one is the well known mass function (relativistic square of momentum). The second one is determined by Pauli - Lubanski four-vector
\begin{eqnarray}
w_\alpha=\frac{1}{2}\varepsilon_{\alpha\beta\gamma\delta}\mathcal{J}_I^{\beta\gamma}p^\delta= \frac{1}{2}\varepsilon_{\alpha\beta\gamma\delta}S_I^{\beta\gamma}p^\delta. \label{Pauli_Lubanski}
\end{eqnarray}
The invariant square of the above vector can be easily calculated in the rest frame. It is not difficult to show that in the frame with the beginning in the mass centre ($\vec{\mathrm{x}}_c=0$) one has
\begin{eqnarray}
\mathcal{J}_I^{0i}\rightarrow\mathrm{J}_I^{0i}= \mathrm{S}_I^{0i}=0,\quad 
\mathrm{x}^i= \frac{1}{2m(p)}\pi_I^\alpha[\gamma^{0i}]_{\alpha\beta}\eta_I^\beta.
\label{pauli_l_passage}
\end{eqnarray}
The components of Pauli- Lubanski vector in the mass centre frame are given by 
\begin{eqnarray}
\mathfrak{w}^0=0,\quad \vec{\mathfrak{w}}=-m(p)\vec{\mathrm{S}}_I,\quad w^\mu w_\mu=\mathfrak{w}^\mu \mathfrak{w}_\mu=m(p)^2 \mathrm{S}_I^2,
\label{p-l components}
\end{eqnarray}
which clearly indicates that the quantity $\mathrm{S}_I$ is Poincare invariant and this fact justifies its name "spin".
The present considerations are summarized by the analysis of the set of vectors
$\dot{\vec{\mathrm{x}}}_I$, $\vec{\mathrm{S}}_I$, $\vec{\mathrm{x}}_I\;$:
\begin{eqnarray}
\dot{\mathrm{x}}_I^{i}&=(-1)^I\frac{1}{2}h\eta_I^{(0)\alpha}[\gamma^i]_{\alpha\beta}\eta_I^{(0)\beta},\label{v.1}\\
\mathrm{x}_I^{i}&=-\frac{1}{2}\frac{g}{m(p)}\eta_I^{(0)\alpha}[\gamma^{0i}]_{\alpha\beta}\eta_I^{(0)\beta},\label{v.2}\\
\mathrm{S}_I^{i}&=-\frac{1}{4}g\varepsilon_{ijk}\eta_I^{(0)\alpha}[\gamma^{jk}]_{\alpha\beta}\eta_I^{(0)\beta},
\label{set}
\end{eqnarray}
expressed in the rest frame. These vectors form the orthogonal frame (fig.\ref{fig.1.3}) and in addition:
\begin{eqnarray}
\frac{\vec{\mathrm{S}}_I}{|\vec{\mathrm{S}}_I|}=(-1)^I\frac{\dot{\vec{\mathrm{x}}}_I}{|\vec{\mathrm{x}}_I|}\times \frac{\vec{\mathrm{x}}_I}{|\dot{\vec{\mathrm{x}}}_I|}, \quad |\vec{\mathrm{S}}_I|=m(p)|\vec{\mathrm{x}}_I|=\frac{g}{h}|\dot{\vec{\mathrm{x}}}_I|=\frac{g}{2}\mathrm{j}_I^{0}.
\label{orto_set}
\end{eqnarray}
 The above formulas and pictures indicate the opposite orientations of spin with respect of the product 
$\vec{\mathrm{x}}\times\dot{\vec{\mathrm{x}}}$ depending on the sign of the couplings of spinors with trajectory. For $I=1,\,\,2$ the orientation of spin is in agreement or opposite. The corresponding spinning particle will be called orbital or anti-orbital correspondingly.\\
The above description of the motion can be boosted to the  frame with non zero spatial momentum of the particle. In such a frame the motion is a spiral spooled on a elliptic cylinder centred along the straight line determined by the momentum. The jump of the spiral 
\begin{eqnarray}
\Delta l=2\pi \frac{h}{\Omega_p}\sqrt{\left(\frac{\mathrm{S}_I}{g}\right)^2+ \left(\frac{m_0}{h}\right)^2}
\label{jump}.
\end{eqnarray}
clearly indicates the dependence of the inertia of the particle (mass) on the spin, as already pointed in the relation (\ref{mass spin}). \\
It should be stressed that despite of the fact that $\dot{x}^0$ contains oscillating terms the coordinate $x^0$ (time) remains the monotonically growing function of the parameter $\tau$. Thanks to this property the value of the speed can be determined from the formula
\begin{eqnarray}
v=|\dot{\vec{x}}|/\dot{x}^0=\sqrt{1-(m_0/\dot{x}^0)^2}
\label{speed 1}.
\end{eqnarray}
\subsection{Zitterbewegung}
The original evolution parameter $\tau$ is not a proper time of the mass centre. Replacing it by the proper time $\tau_p =\mathrm{x}^0$ and making use of the formulae (\ref{x solution}) the velocity $u_I^\mu=d x_I^\mu/d \tau_w$ is defined and evaluated as
\begin{eqnarray}
u_I^\mu=\frac{p^\mu}{m}+\frac{(-1)^I h}{m}\left(\delta^\mu_\nu+\frac{ p^\mu p_\nu}{m^2}\right)\frac{j^\nu_I(0)}{A} \cos \omega_p\tau_p- \frac{2h}{g m}\frac{Z^{\mu\nu}(0)p_\nu}{m A}\sin \omega_p\tau_p,  
\label{velocity}
\end{eqnarray}
where 
\begin{eqnarray}
 A=1+\frac{m_0^2}{m^2},\quad\quad \omega_p=2\frac{\Omega_p}{A m}.\label{definitions}
\end{eqnarray}
It is worth to notice that the standard relation for relativistic velocity 
$u^\mu u_\mu=-1\,$ is satisfied. 
For the orbital ($I = 1$) particle the limit  $m_0\rightarrow 0$ gives
\begin{eqnarray}
m\rightarrow 2\frac{h}{g}\mathrm{S}_1,\quad\quad A\rightarrow 1,\quad\quad \omega_p \rightarrow \frac{m}{\mathrm{S}_1}, \quad\quad \mathfrak{u}_1^\mu \mathfrak{u}_{1\mu}=0.
\label{limit}
\end{eqnarray}
The formula (\ref{velocity}) corresponding to the result \cite{martin} is its generalization  \cite{barut1} (see also \cite{Rivas_1}, \cite{Rodri}) for the particles with arbitrary spin \footnote{The consistent description leading to the dependence of this type was firt given by  Yakov Frenkel in 1926 r \cite{Kosyakov}.}:
\begin{eqnarray}
\mathfrak{u}_1^\mu=\frac{p^\mu}{m}+\left(\mathfrak{u}_1^\mu(0)-\frac{p^\mu}{m}\right) \cos \frac{m}{\mathrm{S}_1}\tau_p- \frac{1}{\mathrm{S}_1 m}Z^{\mu\nu}(0)p_\nu\sin \frac{m}{\mathrm{S}_1}\tau_p.  
\label{velocity Zitter}
\end{eqnarray}
In the centre of mass frame the particle circles around the mass centre with the speed of light 
($|\vec{\mathfrak{u}}_1| \rightarrow 1$ and frequency  $\omega_p$ being in inverse proportion to the value of spin. For the spin of electron  $\mathrm{S}_2=1/2$ from (\ref{velocity Zitter}) one obtains the formula for position depending on time of the mass centre $\tau_c=\tau_p p^0/m$:
\begin{eqnarray}
x^i_1=\frac{p^i}{p^0}\tau_c+\frac{1}{2m}\left(\mathfrak{u}_1^i(0)-\frac{p^i}{m}\right) \sin \frac{2m^2}{p^0}\tau_c+Z^{i\nu}(0) \frac{p_\nu}{m^2}\cos \frac{2m^2}{p^0}\tau_c. 
\label{class Zitter}
\end{eqnarray}
This motion describes the classical counterpart of the phenomenon of "Zitterbewegung" discovered by Schr\"{o}dinger in the process of analysis of the changes average of the position operator $\hat{\vec{x}}$ of the particle of spin $1/2$ whose motion was governed by Dirac Hamiltonian
\cite{Greiner_rel}, \cite{Thaller1}, \cite{Thaller2}, \cite{barut2}, \cite{Simulik}, \cite{Rivas_2}.
\subsection{Magnetic and dipole moments. PCT transformation.}
 Determination of the gyromagnetic ratio  $g_m=2$ was recognized as a success of Dirac theory of electron for many years. Later on  Levy-Leblond obtained  $g_m =2$ in the case  of non relativistic vawe equation  for particle of   spin $s= 1/2$. Proca  obtained $g_m = 1$ for the particle of spin $1$.  Belinfante supposed that the gyromagnetic ratio for elementary systems of spin $s$ equals to  $g_m = 1/s$ independently of the value of spin. He showed that the formula is true for the systems with spin $3/2$. After few years the supposition was analyzed by  Moldauer and Case. They confirmed the formula  for half integral spins.  Using minimal coupling with electromagnetic field  Tumanov showed this dependence for $s = 2$.\newline
Weinberg predicted  $g_m= 2$  for intermediate  bosons  of weak interactions. His prediction was confirmed by discovery of charged bosons $W^\pm$ of spin $1$ and $g_m= 2$, in contrast with supposition of Belinfante. The fundamental question was raised:  if  $g_m = 2$ for a particle of arbitrary spin? 
\newline
In this section the elementary electromagnetic properties of charged spinning particle are described. The magnetic moment and electric dipole moment are determined. This allows one to try to  determine the gyromagnetic ratio on the classical level.
\newline
It can be assumed that the system consists of spinning particle charged by the electric charge $q$ and 
circulating around mass centre with velocity $\vec{u}_I=\partial_{\tau_p}\vec{\mathrm{x}}_I$.  In consequence the particle generates a magnetic moment:
\begin{eqnarray}
\vec{\mathfrak{M}}_I=\frac{1}{2}\varint\vec{\mathrm{x}}'\times \vec{\mathfrak{J}}_I(\vec{\mathrm{x}}')d^3\vec{\mathrm{x}}',  
\label{magnetic moment}
\end{eqnarray}
where $\vec{\mathfrak{J}}_I(\vec{\mathrm{x}}')= q\vec{u}_I\delta(\vec{\mathrm{x}}'-\vec{\mathrm {x}}_I)$ is a vector of current of circulating charge. The motion is considered with respect to the frame of the mass centre.\\
With a little bit of effort it can be calculated that
\begin{eqnarray}
\vec{\mathfrak{M}}_I=q\frac{(-1)^{I+1}}{2m(p)}g_m\vec{\mathrm{S}}_I,\quad \mathrm{where:}\quad g_m=\frac{h}{g}\frac{\mathrm{S}_I}{\dot{\mathrm{x}}^0}.  
\label{moment mag calc}
\end{eqnarray}
It is easy to see that the above, classical gyromagnetic ratio is bounded: $g_m\leq 1$. The upper bound 
 $g_m=1$ corresponds to the motion with the speed of light. Strong suggestion follows from the above: the attempts to determine correct gyromagnetic ratio on the classical grounds \cite{Rivas_1} are bound to fail.
The moving charge is a source of electric dipole moment:
\begin{eqnarray}
\vec{\mathfrak{d}}_I=q\vec{\mathrm{x}}_I=(-1)^I q\frac{g}{h\mathrm{S}_I} \frac{\dot{\mathrm{x}}^0}{m(p)}\vec{\mathrm{S}}_I\times \vec{u}_I.  
\label{dipol}
\end{eqnarray} 
For the luxon particle ($I=1$) of spin $1/2$ with charge of electron ($q=\mathrm{e}$) one obtains
\begin{eqnarray}
\vec{\mathfrak{M}}_1=\frac{\mathrm{e}}{2m(p)}\vec{\mathrm{S}}_1,\quad\quad \vec{\mathfrak{d}}_1=-\frac{\mathrm{e}}{m(p)}\vec{\mathrm{S}}_1\times \vec{u}_I.  
\label{elektron}
\end{eqnarray}
The single spinor system under consideration is not $PT$ invariant. \\
Introducing operators of space reflection $P=\gamma^0$ and time inversion $T=\gamma^0\gamma^5$ with their natural actions:
\begin{eqnarray}
P(\gamma^0,\vec{\gamma})P^{-1}=(\gamma^0,-\vec{\gamma}),\quad\quad T(\gamma^0,\vec{\gamma})T^{-1}=(-\gamma^0,\vec{\gamma}),  
\label{PT}
\end{eqnarray}
one easily calculates that the Lagrange function transforms as:
\begin{eqnarray}
\mathcal{L}_I\rightarrow{} \mathcal{L}_I^{PT}=\frac{1}{2} \left(e^{-1}g_{\mu\nu}\dot{x}^\mu\dot{x}^\nu-e m_0^2+(-1)^{I} h\dot{x}j_I+2g \bar{\eta}_I\dot{\eta}_I\right).
\label{PT lagrange}
\end{eqnarray}
 This transformation corresponds to a change of the sign of the coupling of spinor with trajectory. It seems that the unique and simplest way to save discreet spacetime symmetries of the system is to supplement them by the transformation $C$ of charge conjugation. In order to define it it is necessary to introduce in the system with two spinorial variables and with opposite couplings with trajectory.
The supplemented Lagrange function
\begin{eqnarray}
\label{PTC_invariant}
\mathcal{L}=\frac{1}{2} \left(e^{-1}g_{\mu\nu}\dot{x}^\mu\dot{x}^\nu-e m_0^2- h\dot{x}j_1+2g \bar{\eta}_1\dot{\eta}_1+h\dot{x}j_2+2g \bar{\eta}_2\dot{\eta}_2\right)
\end{eqnarray} 
is invariant with respect to the combined $CPT$ transformation, provided
\begin{equation}
C\eta_1=\eta_2,\quad\quad C\eta_2=\eta_1.
\label{C transform}
\end{equation}
This property justifies considerations of the system governed by the action corresponding to modified  Lagrange function. Its analysis is performed in the next section.
\section{Nonlinear system with two spinors}
\begin{figure}[ht]
\begin{center}
\includegraphics[width=14.cm]{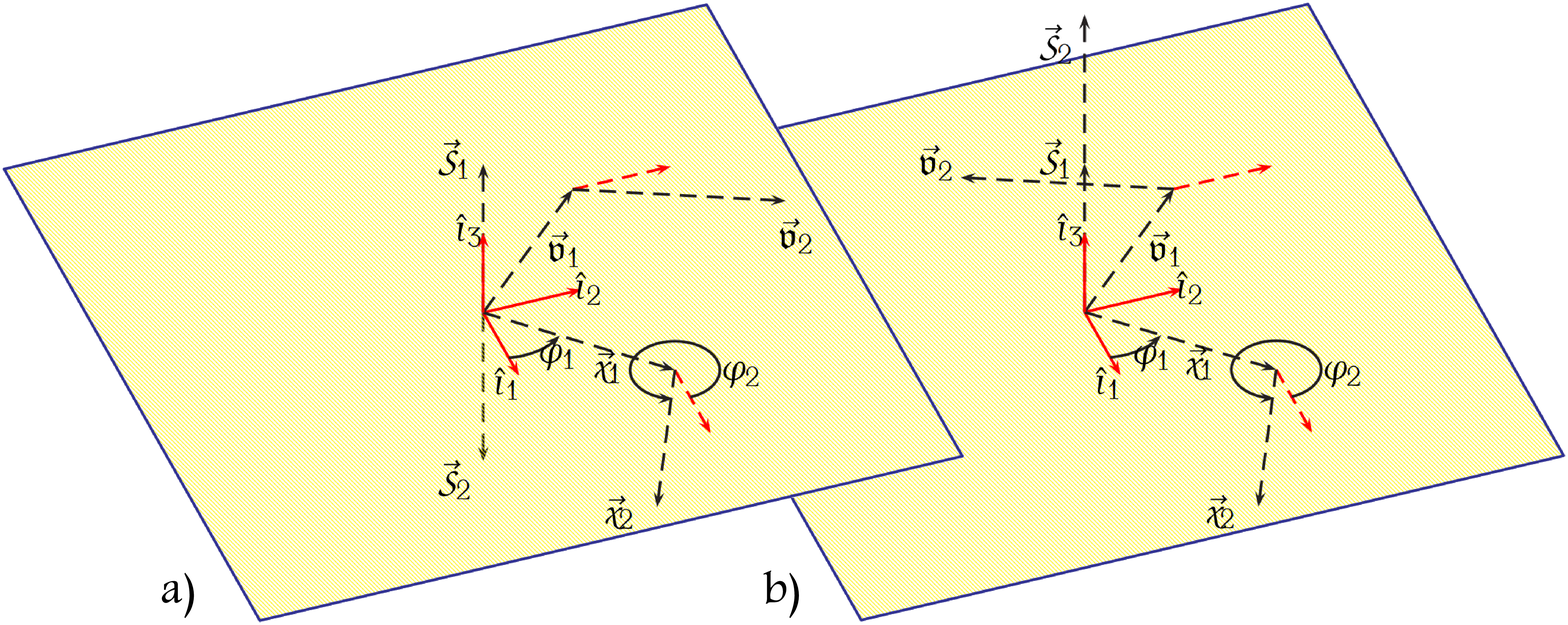}
\caption{Configuration of vectors $\vec{\mathrm{x}}_I$, ${\vec{\mathfrak{v}}}_I$ and $\vec{\mathrm{S}}_I$ for a) $\vec{\mathrm{S}}_2\cdot\hat{\imath}_3=-\mathrm{S}_2 $ b) $\vec{\mathrm{S}}_2\cdot\hat{\imath}_3=\mathrm{S}_2 $.}
\label{fig.2.1.2}
\end{center}
\end{figure}

According to what was said in the last section the extended system containing two spinors will be analyzed in this chapter. The corresponding Lagrange function is defined to be 
\begin{eqnarray}
\mathcal{L}=-\frac{1}{2}(e^{-1}\dot{x}^2-em_0^2)+\sum_{I\in\left\{1,2\right\}}\mathcal{L}_I,
\label{lagrange 2}
\end{eqnarray}
where $\mathcal{L}_I$ is already defined in (\ref{lagrange spinor}). The canonical momenta $\pi_e$ and $\pi_{I\alpha}$ conjugated to the variables  $e$ and  $\eta_I^\alpha$ are expressed by the formulas (\ref{momentum e}) and (\ref{momentum spin}). The momentum $p_\mu$ conjugated to the position coordinates $x^\mu$ contains now the contributions of two spinors:
\begin{eqnarray}
p_\mu=\frac{\partial\mathcal{L}_n}{\partial \dot{x}^\mu}=e^{-1}\dot{x}_{\mu}-\frac{h}{2}\sum_I (-1)^I \mathrm{j}_{I\mu}.
\label{x momentum 2}
\end{eqnarray}
Adopting standard gauge  $e=1$ and using Lagrange equations on obtains the same restriction on the derivatives of position as in the case of single spinor:
\begin{eqnarray}
\dot{x}^2=-m_0^2.
\label{restriction mass}
\end{eqnarray}
The equations of motion generated by (\ref{lagrange 2}) 
\begin{eqnarray}
\dot{x}^\mu&=&p^\mu+\frac{h}{2}\sum_I (-1)^I \mathrm{j}_I^\mu, \label{x dot}\\
g\dot{\eta}_{1\alpha} &=&-\frac{h}{2}p_\mu[C\gamma^\mu]_{\alpha\beta}\eta_1^\beta -\frac{h^2}{4}\mathrm{j}_{2\mu}[C\gamma^\mu]_{\alpha\beta}\eta_1^\beta, \label{eta one dot}\\
g\dot{\eta}_{2\alpha} &=&\frac{h}{2}p_\mu[C\gamma^\mu]_{\alpha\beta}\eta_2^\beta -\frac{h^2}{4}\mathrm{j}_{1\mu}[C\gamma^\mu]_{\alpha\beta}\eta_2^\beta, \label{eta two dot}
\end{eqnarray}
are essentially different from those corresponding to  single spinor system. The formulas (\ref{eta one dot}) and (\ref{eta two dot}) define a non linear system of equations corresponding to mutual interaction of spinors. This is the price for CPT invariance of the system, which complicates the search for general solution drastically.\\
Nevertheless one can make use some simplifications by exploiting conservation laws and using the rest frame ($\vec{p}=0$).
Lorenz invariance of the system implies (via Noether theorem) six constants of motion:
\begin{eqnarray}
\mathrm{J}^{0i}&=&-m\mathrm{x}^i -\frac{g}{2}\sum_I\eta_I^{(0)\alpha}[\gamma^{0i}]_{\alpha\beta}\eta_I^{(0)\beta},
\label{J0i}\\
\mathrm{S}^{ij}&=&-\frac{g}{2}\sum_I\eta_I^{(0)\alpha}[\gamma^{ij}]_{\alpha\beta}\eta_I^{(0)\beta}.
\label{Sij}
\end{eqnarray}
Now, one can choose the beginning of the coordinate frame in such a way that  $\mathrm{J}^{0i}=0$. One can also define the spin vector:$\mathrm{S}_i=\varepsilon_{ijk}\mathrm{S}^{jk}/2$. Applying formulas (\ref{x momentum 2}), (\ref{J0i}) and (\ref{Sij}) one obtains a set of vectors
\begin{eqnarray}
\vec{\mathrm{x}}=\sum_I \vec{\mathrm{x}}_I,\quad \dot{\vec{\mathrm{x}}}=\sum_I \vec{\mathfrak{v}}_I, \quad \mathrm{const}=\vec{\mathrm{S}}=\sum_I \vec{\mathrm{S}}_I,\label{set vect}
\end{eqnarray}
where
\begin{eqnarray}
\mathrm{x}_I^{i}&=&-\frac{1}{2}\frac{g}{m}\eta_I^{(0)\alpha}[\gamma^{0i}]_{\alpha\beta}\eta_I^{(0)\beta}, \label{x 2}\\
\mathfrak{v}_I^{i}&=&(-1)^I\frac{1}{2}h\eta_I^{(0)\alpha}[\gamma^i]_{\alpha\beta}\eta_I^{(0)\beta}, \label{v 2}\\
\mathrm{S}_I^{i}&=&-\frac{1}{4}g\varepsilon_{ijk}\eta_I^{(0)\alpha}[\gamma^{jk}]_{\alpha\beta}\eta_I^{(0)\beta}. \label{S 2}
\end{eqnarray}
The vectors above do form two orthogonal bases ($I\in\{1,2\}$):
\begin{eqnarray}
(-1)^I\frac{\vec{\mathfrak{v}}_I}{|\vec{\mathfrak{v}}_I|}= \frac{\vec{\mathrm{x}}_I}{|\vec{\mathrm{x}}_I|}\times \frac{\vec{\mathrm{S}}_I}{|\vec{\mathrm{S}}_I|}, \quad\quad 0=\vec{\mathrm{x}}_I\cdot\vec{\mathfrak{v}}_I=\vec{\mathrm{x}}_I\cdot\vec{\mathrm{S}}_I= \vec{\mathfrak{v}}_I\cdot\vec{\mathrm{S}}_I. 
\label{ortho}
\end{eqnarray}
In contrast to the system with single spinor a vector  $\vec{\mathfrak{v}}_I$ is not the derivative of the position vector $\vec{\mathrm{x}}_I$ corresponding to $I$-th component. From (\ref{x 2}) and the equations of motion (\ref{eta one dot} - \ref{eta two dot}) it follows that
\begin{eqnarray}
\dot{\vec{\mathrm{x}}}_I-\vec{\mathfrak{v}}_I=-(-1)^I \frac{h^2}{g^2 m}\mathrm{S}_1\mathrm{S}_2\sum_J \frac{\vec{\mathfrak{v}}_J}{|\vec{\mathfrak{v}}_J|}, \quad\quad \mathrm{S}_I=m|\vec{\mathrm{x}}_I|=\frac{g}{h}|\vec{\mathfrak{v}}_I|=\frac{g}{2}|\mathrm{j}_I^0|. \label{interest}
\end{eqnarray}
Making use of the definitions (\ref{v 2} - \ref{S 2}), relation (\ref{interest}) and the equations 
(\ref{eta one dot} - \ref{eta two dot}) one can determine the derivatives of the vectors 
$\vec{\mathfrak{v}}_I$ and $\vec{\mathrm{S}}_I$:
\begin{eqnarray}
\dot{\vec{\mathfrak{v}}}_I = (-1)^I\left(\frac{h}{g}\right)^2 m\frac{\dot{\vec{\mathrm{x}}}_I}{|\vec{\mathfrak{v}}_I|}\times \vec{\mathrm{S}}_I, \quad  \quad \dot{\vec{\mathrm{S}}}_I =(-1)^I\vec{\mathfrak{v}}_1\times\vec{\mathfrak{v}}_2. 
\label{derivatives}
\end{eqnarray}
From the above it is evident that the vector $\vec{\mathfrak{v}}_I$ changes in the plane perpendicular 
$\vec{\mathrm{S}}_I$ i.e. spanned by the vectors $\vec{\mathfrak{v}}_I$ and $\vec{\mathrm{x}}_I$. Since 
the systems (\ref{ortho}) are orthogonal the vector $\vec{\mathrm{x}}_I$ should change in the same plane.
The property of $\vec{\mathrm{x}}_I$ i $\vec{\mathrm{S}}_I$ being orthogonal together with the equations (\ref{interest}) enforces the orthogonality of $\vec{\mathrm{S}}_I$ to the plane spanned by the vectors  $\vec{\mathfrak{v}}_J$ i $\vec{\mathrm{x}}_J$ corresponding to accompanying spinor ($J\neq I$). This means that the vectors $\vec{\mathrm{x}}_1$, $\vec{\mathfrak{v}}_1$ and $\vec{\mathrm{x}}_2$, $\vec{\mathfrak{v}}_2$ are located in the same plane. Consequently the vectors $\vec{\mathrm{S}}_1$ and $\vec{\mathrm{S}}_2$ are parallel.\\
The equations (\ref{derivatives}) enable one to determine the derivatives of the spins $\mathrm{S}_I=|\vec{\mathrm{S}}_I|$:
\begin{eqnarray}
\dot{\mathrm{S}}_I=\frac{\vec{\mathrm{S}}_I}{\mathrm{S}_I}\cdot\dot{\vec{\mathrm{S}}}_I= (-1)^I\frac{\vec{\mathrm{S}}_I}{\mathrm{S}_I}\cdot (\vec{\mathfrak{v}}_1\times\vec{\mathfrak{v}}_2),
\nonumber
\end{eqnarray}  
which after use of (\ref{ortho}) and obvious identity gives 
\begin{eqnarray}
\dot{\mathrm{S}}_1=\frac{h m}{g}\vec{\mathrm{x}}_1\cdot\vec{\mathfrak{v}}_2, \quad \quad \dot{\mathrm{S}}_2=-\frac{h m}{g}\vec{\mathrm{x}}_2\cdot\vec{\mathfrak{v}}_1. 
\label{spinors 2}
\end{eqnarray}
The total spin $\vec{\mathrm{S}}$  of the resulting system is conserved (\ref{set}) and the contributions $\vec{\mathrm{S}}_1$ i $\vec{\mathrm{S}}_2$ of the individual spinors are parallel. Consequently the motion of the particle (in the rest frame) must be flat.\\
In order to push forward the calculations it is convenient to introduce right-hand orthonormal frame consisting of orthonormal basis vectors $\hat{\imath}_3=\vec{\mathrm{S}}_1/|\vec{\mathrm{S}}_1|$ and two complementary $\hat{\imath}_1$, $\hat{\imath}_2$ chosen in such a way that  $\hat{\imath}_1\times\hat{\imath}_2=\hat{\imath}_3$.\\
It is now natural to introduce  polar coordinates $\mathrm{S}_I$ i $\varphi_I$ to investigate the motion of spinning system. In such coordinates one can write:
\begin{eqnarray}
\vec{\mathrm{x}}_I &=&\frac{\mathrm{S}_I}{m}(\cos\varphi_I\hat{\imath}_1+\sin\varphi_I\hat{\imath}_2), \label{x polar}\\
\vec{\mathfrak{v}}_I &=&-(-1)^I \frac{h}{g}\vec{\mathrm{S}}_I\cdot \hat{\imath}_3(-\sin\varphi_I\hat{\imath}_1+\cos\varphi_I\hat{\imath}_2). \label{v polar}
\end{eqnarray}
Introducing new basis vectors
\begin{eqnarray}
\hat{\mathfrak{e}}_{\mathfrak{r}_I}=\vec{\mathrm{x}}_I/|\mathrm{x}_I|,\quad \hat{\mathfrak{e}}_{\varphi_I}=\frac{d}{d\varphi_I}\frac{\vec{\mathrm{x}}_I}{|\mathrm{x}_I|},\quad \mathrm{where}:\;\;|\mathrm{x}_I|=\mathrm{S}_I/m
\label{new basis}
\end{eqnarray}
one can rewrite (\ref{x polar} - \ref{v polar}) in the form:
\begin{eqnarray}
\vec{\mathrm{x}}_I=\frac{\mathrm{S}_I}{m}\hat{\mathfrak{e}}_{\mathfrak{r}_I}, \quad \vec{\mathfrak{v}}_I &=-(-1)^I \frac{h}{g}(\vec{\mathrm{S}}_I\cdot \hat{\imath}_3) \hat{\mathfrak{e}}_{\varphi_I}.
\label{new polar}
\end{eqnarray}
Using the above equations one can find the expression for angular velocities
\begin{eqnarray}
\dot{\varphi}_I=-(-1)^I\frac{g m}{h}\frac{\vec{\mathrm{S}}_I\cdot \hat{\imath}_3}{\mathrm{S}_I}\frac{\vec{\mathfrak{v}}_I\cdot \dot{\vec{\mathrm{x}}}_I}{\mathrm{S}_I^2}. \label{angular velocity}
\end{eqnarray}
and taking into account (\ref{interest}): 
\begin{eqnarray}
\frac{\vec{\mathrm{S}}_I\cdot\hat{\imath}_3}{\mathrm{S}_I}\dot{\varphi}_I &=-(-1)^I\frac{h m}{g}+ \left(\frac{h}{g}\right)^2 \frac{\mathrm{S}_1\mathrm{S}_2}{\mathrm{S}_I}+ \frac{\vec{\mathfrak{v}}_1\cdot \vec{\mathfrak{v}}_2}{\mathrm{S}_I}.
\label{angular final}
\end{eqnarray}
The equations (\ref{spinors 2}) and (\ref{angular final}) can be rewritten in polar coordinates 
$\mathrm{S}_I$ i $\varphi_I$:
\begin{eqnarray}
\left\{\;\dot{\mathrm{S}}_1\;,\;\dot{\mathrm{S}}_2\;\right\} &=&\left(\frac{h}{g}\right)^2\mathrm{S}_1\mathrm{S}_2 \left\{\;\frac{\vec{\mathrm{S}}_2\cdot\hat{\imath}_3}{\mathrm{S}_2}\;,\;-1\;\right\} \sin(\varphi_2-\varphi_1), \label{eq trig}\\
\dot\varphi_1 &=&\frac{h m}{g}+\left(\frac{h}{g}\right)^2 \mathrm{S}_2 \left(1-\frac{\vec{\mathrm{S}}_2\cdot\hat{\imath}_3}{\mathrm{S}_2}\cos(\varphi_2-\varphi_1)\right), \label{eq trig 2}\\
\frac{\vec{\mathrm{S}}_2\cdot\hat{\imath}_3}{\mathrm{S}_2} \dot\varphi_2 &=&-\frac{h m}{g}+\left(\frac{h}{g}\right)^2 \mathrm{S}_1 \left(1-\frac{\vec{\mathrm{S}}_2\cdot\hat{\imath}_3}{\mathrm{S}_2}\cos(\varphi_2-\varphi_1)\right), \label{eq trig 3}  
\end{eqnarray}
The above equations together with constraint(\ref{restriction mass}):
\begin{eqnarray}
\frac{m^2-m_0^2}{\mathrm{S}_1 \mathrm{S}_2} =2\frac{h m}{g}\left(\frac{1}{\mathrm{S}_2}-\frac{1}{\mathrm{S}_1}\right)+2\left(\frac{h}{g}\right)^2 \left(1-\frac{\vec{\mathrm{S}}_2\cdot\hat{\imath}_3}{\mathrm{S}_2}\cos(\varphi_2-\varphi_1)\right), \label{ang restriction}  
\end{eqnarray}
describe the dynamics of the particle completely.\\
The solutions of the equations depend on mutual orientation of spins of the spinorial ingredients. For this reason it is necessary to consider two cases $\vec{\mathrm{S}}_2\cdot\hat{\imath}_3=\pm\mathrm{S}_2$ (fig.\ref{fig.2.1.2}) with the equations (\ref{eq trig} - \ref{ang restriction}) rewritten accordingly.

\subsection{Configuration $\vec{\mathrm{S}}_1\cdot\vec{\mathrm{S}}_2=-\mathrm{S}_1\mathrm{S}_2 $ }
It is convenient to introduce new variable $\phi=\varphi_1-\varphi_2$ and to convert the system of equations (\ref{eq trig} - \ref{eq trig 3}) to take the form:
\begin{eqnarray}
\dot{\mathrm{S}}_1 &=&\dot{\mathrm{S}}_2 =\left(\frac{h}{g}\right)^2\mathrm{S}_1\mathrm{S}_2 \sin\phi,\nonumber\\  
\dot\phi &=&\left(\frac{h}{g}\right)^2 (\mathrm{S}_1+\mathrm{S}_2) \left(1+\cos\phi\right), 
\nonumber \\
\dot\varphi_2+\dot\phi &=&\frac{h m}{g}+\left(\frac{h}{g}\right)^2 \mathrm{S}_2 \left(1+\cos\phi\right). 
\label{eq for spins 1}
\end{eqnarray}
From the above it is evident that the variable $\phi$ is non decreasing function of $\tau$. The constraint equation (\ref{ang restriction}):
\begin{eqnarray}
m^2-m_0^2&=2\frac{h m}{g}(\mathrm{S}_1-\mathrm{S}_2)+2\left(\frac{h}{g}\right)^2 \mathrm{S}_1\mathrm{S}_2\left(1+\cos\phi\right), 
\end{eqnarray}
fixes the value of $\phi$. The constraint equation can be satisfied for $\phi = \pi$ only.\\
The resulting system describes the motion corresponding to orbital or antiorbital particle depending on the sign of the difference $\mathrm{S}_1-\mathrm{S}_2$. The spin of this particle equals to  $\mathrm{S}=|\mathrm{S}_1-\mathrm{S}_2|$ and its mass is given by the relation:
\begin{eqnarray}
0=m^2-m_0^2-2\frac{h m}{g}\mathrm{sign}(\mathrm{S}_1-\mathrm{S}_2)\mathrm{S}. 
\label{spin mass relation 1}
\end{eqnarray}
This kind of motion was discussed  in the previous section.  
\subsection{Configuration  $\vec{\mathrm{S}}_1\cdot\vec{\mathrm{S}}_2=\mathrm{S}_1\mathrm{S}_2 $.}
\begin{figure}[ht]
\begin{center}
\includegraphics[width=14.cm]{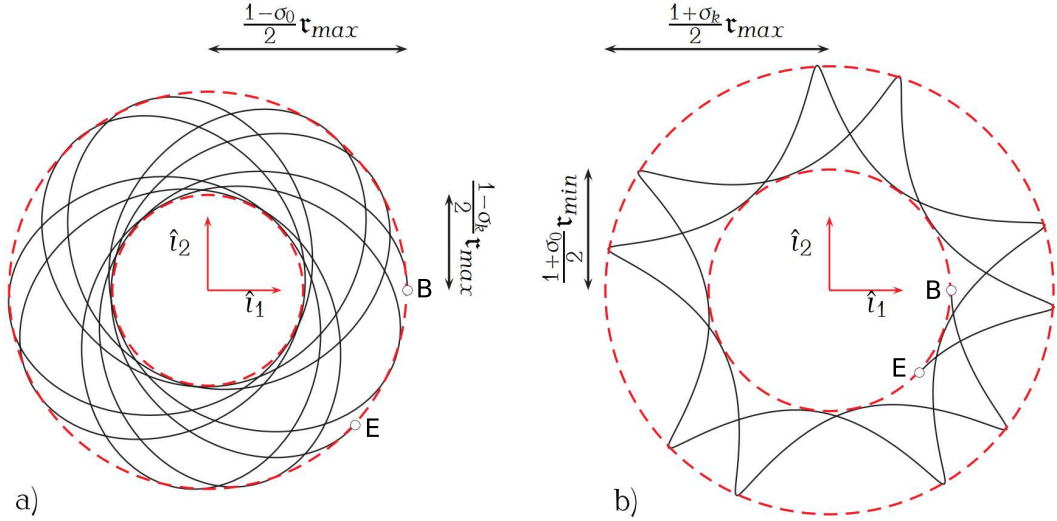}
\caption{The graphs a) $\vec{\mathrm{x}}_1$  b) $\vec{\mathrm{x}}_2$ for $\sigma_k= 0.4$, $m_0=h=g=1$, $S=3$, $\mathfrak{r}_{max}=\mathrm{S}/m$ and from range $\phi\in\langle \phi_{B},\phi_{E}\rangle=\langle 0,20\pi\rangle$.}
\label{fig.2.3}
\end{center}
\end{figure}

\begin{figure}[ht]
\begin{center}
\includegraphics[width=14.cm]{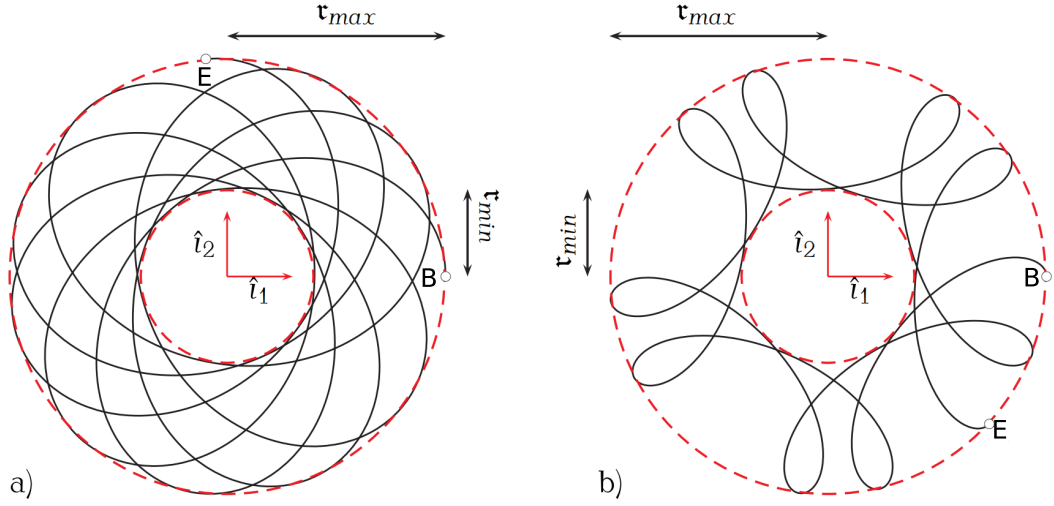}
\caption{The graphs for $\vec{\mathrm{x}}$ for a)  $\sigma_k=-0.4$, b) $\sigma_k=0.4$, corresponding to 
 the range $\phi\in\langle \phi_{B},\phi_{E}\rangle=\langle 0,20\pi\rangle$ ($m_0=h=g=1$, $S=3$)}
\label{fig.limits}
\end{center}
\end{figure}

\begin{figure}[ht]
\begin{center}
\includegraphics[width=14.cm]{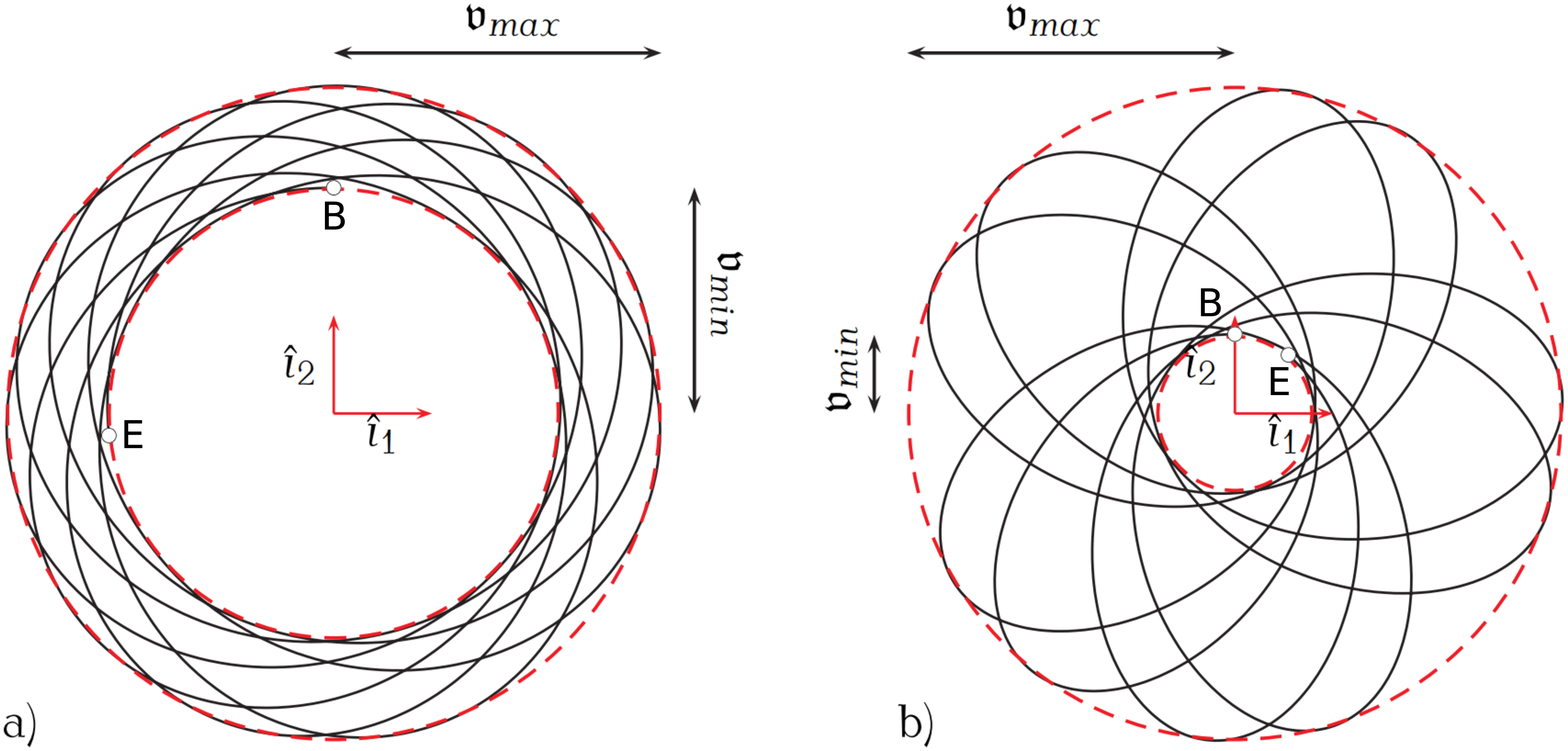}
\caption{The graphs of  $\dot{\vec{\mathrm{x}}}$ for a) $(\sigma_k,\sigma_0)=-(0.4,0.68)$,
 b) $(\sigma_k,\sigma_0)=(0.4,-0.24)$ corresponding to the range  
$\phi\in\langle \phi_{B},\phi_{E}\rangle=\langle 0,20\pi\rangle$ ($m_0=h=g=1$, $S=3$).}
\label{fig.1.10}
\end{center}
\end{figure}

\begin{figure}[ht]
\begin{center}
\includegraphics[width=14.cm]{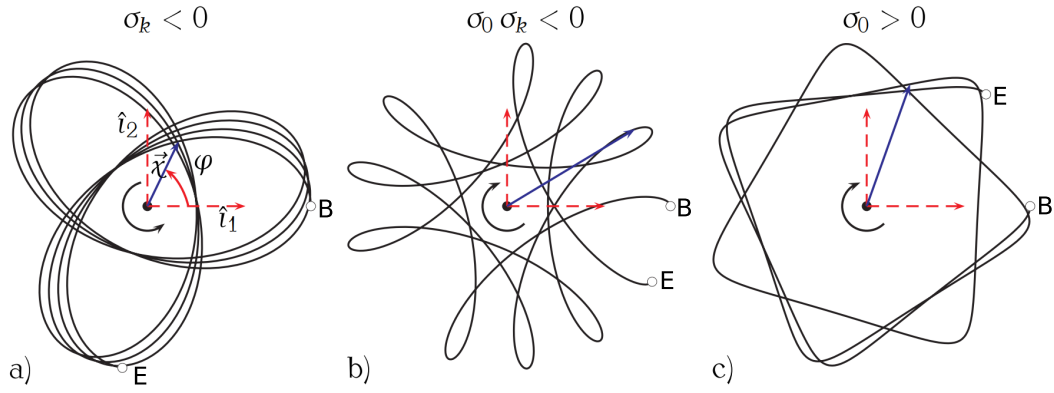}
\caption{Three types of trajectories corresponding to 
evolution $\phi\in(\phi_B,\phi_E)=(0,20\pi)$, for spin $\mathrm{S}=1$ and 
a) $\sigma_k=-0.3$ ($\varphi$ is growing function of $\tau$),
 b) $\sigma_k=0.25$ ($\varphi$ is not monotonic $\tau$),
 c) $\sigma_k=0.68$ ($\varphi$ is decreasing function of $\tau$).}
\label{fig.1.11}
\end{center}
\end{figure}

\begin{figure}[ht]
\begin{center}
\includegraphics[width=14.cm]{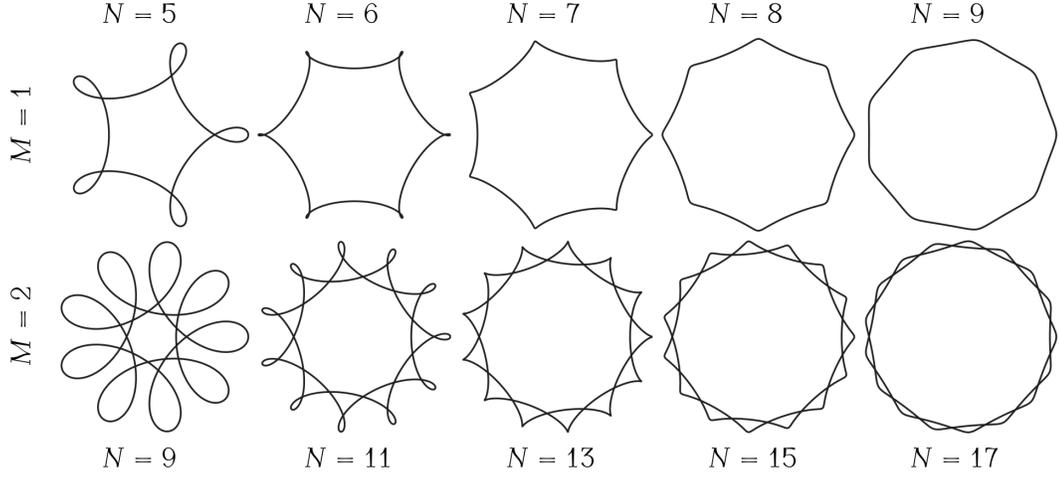}
\caption{Closed trajectories for spin $\mathrm{S}=7/2$ and values of  $\sigma_k$ tables \ref{tab:1} and \ref{tab:2}}
\label{nice picture}
\end{center}
\end{figure}

For such a configuration the equations (\ref{eq trig} - \ref{eq trig 3}) take the following form
\begin{eqnarray}
\dot{\mathrm{S}}_1&=&-\dot{\mathrm{S}}_2= \left(\frac{h}{g}\right)^2\mathrm{S}_1\mathrm{S}_2  \sin(\varphi_2-\varphi_1), \label{new eq 1}\\
\dot\varphi_{1/2} &=&\pm\frac{h m}{g}+\left(\frac{h}{g}\right)^2 \mathrm{S}_{2/1} \left(1-\cos(\varphi_2-\varphi_1)\right), \label{new eq 2}  
\end{eqnarray}
Since the total spin of the system is the sum of contributions from particular spinor components:
$\mathrm{S}=\mathrm{S}_1+\mathrm{S}_2$ it is more convenient to use new variables $\phi$ and $\sigma$:
\begin{eqnarray}
\phi=\varphi_1-\varphi_2,\quad\quad \sigma=\frac{\mathrm{S}_2-\mathrm{S}_1}{\mathrm{S}_1+\mathrm{S}_2}
\label{new var}
\end{eqnarray}
The quantities (\ref{new var}) do satisfy the equations
\begin{eqnarray}
\dot{\sigma} &=&\frac{1}{2}\left(\frac{h}{g}\right)^2 \mathrm{S}(1-\sigma^2) \sin\phi,
\label{sigma dot}\\
\dot\phi &=&2\frac{h m}{g}+2\left(\frac{h}{g}\right)^2 \mathrm{S}\sigma \sin^2\frac{\phi}{2},
\label{phi dot}\\
m^2&=&m_0^2-2\mathrm{S}\frac{h m}{g}\sigma+\mathrm{S}^2\left(\frac{h}{g}\right)^2(1-\sigma^2) \sin^2\frac{\phi}{2}. 
\label{phi sigma const} 
\end{eqnarray}
The last equation (in fact constraint equation) enables one to determine  $\sigma$ as unique function of variable $\phi$:
\begin{eqnarray}
\sigma(\phi)=-\frac{g m}{h \mathrm{S} \sin^2\frac{\phi}{2}}\left[\,1- \sqrt{1-\left(1-\frac{m_0^2}{m^2}-\frac{h^2}{g^2}\frac{\mathrm{S}^2}{m^2} \sin^2\frac{\phi}{2}\right)\sin^2\frac{\phi}{2}}\;\right].
\label{sigma of phi}
\end{eqnarray} 
The function (\ref{sigma of phi}) has two types of turning points 
 $\sigma_0$ (for $\phi=2n\pi$, where $n\in\mathbb{N}$) and $\sigma_k$ (for $\phi=(2n+1)\pi$) which are correlated by 
\begin{eqnarray}
\sigma_0=\sigma_k-\frac{1}{2}\frac{1-\sigma_k^2}{\sqrt{1+\frac{m_0^2 g^2}{h^2 \mathrm{S}^2}}-\sigma_k}.
\end{eqnarray} 
The turning points are extremal points for position vectors $\mathfrak{r}_I=|\mathrm{x}_I|$ of spinorial ingredients
\begin{eqnarray}
\mathfrak{r}_I^{(k)}=\frac{1}{2m}\left[ 1+(-1)^I\sigma_k\right] ,\quad\quad \mathfrak{r}_I^{(0)}=\frac{1}{2m}\left[1+(-1)^I\sigma_0\right],\label{turning ext}
\end{eqnarray}
as well as they are extremal for derivatives  $\dot{\phi}$, $\dot{\varphi}_1$, $\dot{\varphi}_2$:
\begin{eqnarray}
\dot{\phi}^{(0)}&=&\dot{\phi}(2n\pi)=\frac{2h}{g}m,\quad \dot{\phi}^{(k)}=\dot{\phi}([2n+1]\pi)=\frac{2h}{g}m + \frac{2h^2}{g^2}\mathrm{S}\sigma_k,
\nonumber\\
\dot{\varphi}^{(0)}_1 &=&\dot{\varphi}_1(2n\pi)=\frac{h}{g}m,\quad \dot{\varphi}_1^{(k)}=\dot{\varphi}_1([2n+1]\pi)=\frac{h}{g}m + \frac{h^2}{g^2}\mathrm{S}\left[1+\sigma_k\right],
\nonumber\\
\dot{\varphi}^{(0)}_2 &=&\dot{\varphi}_2(2n\pi)=-\frac{h}{g}m,\quad \dot{\varphi}_2^{(k)}=\dot{\varphi}_2([2n+1]\pi)=-\frac{h}{g}m + \frac{h^2}{g^2}\mathrm{S}\left[1-\sigma_k\right].
\label{ekstrema}
\end{eqnarray}
The derivatives of $\phi$, $\varphi_1$, $\varphi_2$ are of constant sign, hence these functions are monotonic in the original parameter $\tau$. Therefore the chronology of events can be controlled by 
the variable $\phi$ to obtain from (\ref{angular velocity}) the following formula:
\begin{eqnarray}
\varphi_{1/2}(\phi)=\pm \frac{1}{2}\phi+\mathcal{C}(\phi)\quad\mathrm{where:}\quad \mathcal{C}(\phi)=\frac{h}{2g}\mathrm{S}\int_0^\phi\frac{sin^2\phi '/2}{m +\frac{h}{g}\mathrm{S}\sigma(\phi ')\sin^2\phi '/2}d\phi '.
\label{katy:varphi}
\end{eqnarray}
The function $\mathcal{C}$ can be expressed in terms of well known  elliptic integrals.
The nonlinear problem is solved analytically thereby.\\
The evolution of vectors $\vec{\mathrm{x}}_I$ can be obtained by the use of (\ref{x polar}).
It is illustrated on figure \ref{fig.2.3}.
It is worth to mention that there are the values of $\sigma_k$ for which the vectors $\vec{\mathrm{x}}_I$ describe closed trajectories. The closed orbits appear for the same values of 
$\sigma_k$. The closed orbits can be determined by finding all simplified fractions $N/M$ satisfying the following relations
\begin{eqnarray}
1\leq\frac{N}{M}=\left\{
\begin{array}{lcl}
2\pi/|\varphi_1(2\pi)|\;\; &\, &\mathrm{for}\;\;  \sigma_k<0\\
2\pi/|\varphi_2(2\pi)|\;\; &\, &\mathrm{for}\;\;  \sigma_k>0\\
\end{array}
\right. , \mathrm{where:}\,\, N,\,M\in \mathbb{N}.\label{ch2:zam} 
\end{eqnarray}
The complete system is described by the sum $\vec{\mathrm{x}}=\vec{\mathrm{x}}_1+\vec{\mathrm{x}}_2$ of the position vectors of the components and their total velocity. The norms of those quantities $\mathfrak{r}=|\vec{\mathrm{x}}|$, $\mathfrak{v}=|\vec{\mathrm{v}}|$ 
\begin{eqnarray}
\mathfrak{r}= \frac{\mathrm{S}}{m}\sqrt{1-(1-\sigma^2)\sin^2\frac{\phi}{2}},\quad\quad \mathfrak{v}= \frac{h\mathrm{S}}{g}\sqrt{1-(1-\sigma^2)\cos^2\frac{\phi}{2}}, 
\label{norms}
\end{eqnarray}
take for $\sigma=\sigma_0$ ($\sigma=\sigma_k$) maximal value  $\mathfrak{r}_{max}$, minimal value $\mathfrak{v}_{min}$ and (minimal value $\mathfrak{r}_{min}$, maximal value $\mathfrak{v}_{max}$) respectively:
\begin{eqnarray}
\mathfrak{r}_{max}&=&\frac{\mathrm{S}}{m},\quad\quad \mathfrak{r}_{min} =\frac{\mathrm{S}}{m}|\sigma_k|, \label{extremal values}\\
\mathfrak{v}_{max}&=&\frac{h\mathrm{S}}{g},\quad\quad \mathfrak{v}_{min}=\frac{h\mathrm{S}}{g}|\sigma_0|.
\end{eqnarray}
It is worth to mention that $\mathfrak{r}_{min}$ and $\mathfrak{r}_{max}$ the radiuses of the concentric circles limiting the particle trajectory (fig.\ref {fig.limits}). By analogy $\mathfrak{v}_{min}$ i $\mathfrak{v}_{max}$ are the limits of the appropriate ring in the space of velocities (fig.\ref {fig.1.10}).
For the parameters  $\sigma_k$ satisfying (\ref{ch2:zam}) the orbits of the complete system are periodic too.
The corresponding $\sigma_k$ can be determined by projecting  $\dot{\vec{\mathrm{x}}}$ onto radial direction and perpendicular one:
\begin{eqnarray}
\dot{\vec{\mathrm{x}}}=\left(\frac{\vec{\mathrm{x}}}{\mathfrak{r}}\cdot \dot{\vec{\mathrm{x}}}\right) \frac{\vec{\mathrm{x}}}{\mathfrak{r}}+\vec{\omega}\times \vec{\mathrm{x}},\quad \mathrm{where}:\; \vec{\omega}=\frac{1}{\mathfrak{r}}\left(\frac{\vec{\mathrm{x}}}{\mathfrak{r}} \times \dot{\vec{\mathrm{x}}}\right).
\label{v projection}
\end{eqnarray}
It is now possible to determine the changes of angular velocity $\dot{\varphi}$ of the particle running around the beginning of the coordinate system:
\begin{eqnarray}
\dot{\varphi}=\vec{\omega}\cdot\hat{\imath}_3= -\frac{h}{g}\frac{\mathrm{S}^2}{m}\frac{\sigma}{\mathrm{r}^2}. 
\label{go around} 
\end{eqnarray}
Depending on the sign of $\sigma$ the particle circles clockwise ($\sigma>0$) or anti-clockwise ($\sigma<0$).\\
It is possible to distinguish three qualitatively different types (phases) of the dynamics of the system:
\begin{itemize}
\item for $\sigma_k<0$, $\varphi$ is  growing function of  $\tau$ (fig. \ref{fig.1.11}a), 
\item for $\sigma_0\,\sigma_k<0$, $\varphi$ is not monotonic function of  $\tau$ (fig. \ref{fig.1.11}b),
\item for $\sigma_0>0$, $\varphi$ is decreasing function of  $\tau$ (fig. \ref{fig.1.11}c).
\end{itemize}
In the case of periodic motion the points of maximal distance from the centre are the apexes of regular N-polygon.
The periodic orbits of N-type closing after M periods around the centre (N is not divisible by M) are obtained for $\sigma_k$ satisfying the condition:

\begin{eqnarray}
1 <\frac{N}{M}=\frac{2\pi}{|\int_0^{T}\dot{\varphi}d\tau|},
\label{condition}
\end{eqnarray}
where $T$ is a period of changes of $\sigma$ (and of $\phi$ and $\mathfrak{r}$) with respect to  $\tau$. Despite of the fact that the explicit dependence of $\varphi$ on $\tau$ is not known it is possible to perform the integration with respect to $\phi$. Using (\ref{go around}) it is possible to obtain
\begin{eqnarray}
\frac{N}{M}=\frac{2\pi}{|\int_0^{2\pi}\frac{\dot{\varphi}}{\dot{\phi}}d\phi|}= \frac{2\pi}{\frac{h}{g} \frac{\mathrm{S}^2}{m}|\int_0^{2\pi}\frac{1}{\mathrm{r}^2}\frac{\sigma}{\dot{\phi}}d\phi|}, 
\label{orbity}
\end{eqnarray}
what is equivalent to (\ref{ch2:zam}).
The illustrative values of $\sigma_k$ for $M=1$ and $M=2$ determined by numerical method are contained in the tables (2.1) and (2.2). The corresponding trajectories are contained in fig.\ref{nice picture}.
\begin{table}[t]
\centering
\begin{tabular}{c|c|c|c|c|c|c|}
\cline{2-7}
$\;$&
$\mathrm{S}=1$&
$\mathrm{S}=3/2$& $\mathrm{S}=2$&$\mathrm{S}=5/2$&$\mathrm{S}=3$&$\mathrm{S}=7/2$\\ \hline
\multicolumn{1}{|c|}{$N=4$}&--&0.5&0.2&0.05 &--&--\\ \hline
\multicolumn{1}{|c|}{$N=5$}&--&--&0.73&0.58&0.5&0.45\\ \hline
\multicolumn{1}{|c|}{$N=6$}&--&--&0.97&0.82&0.74&0.69\\ \hline
\multicolumn{1}{|c|}{$N=7$}&--&--&--&0.95&0.87&0.82 \\ \hline
\multicolumn{1}{|c|}{$N=8$}&--&--&--&--&0.95&0.9\\ \hline
\multicolumn{1}{|c|}{$N=9$}&--&--&--&--&--&0.95 \\ \hline
\multicolumn{1}{|c|}{$N=10$}&--&--&--&--&--&0.99\\ \hline
\end{tabular}
\caption{Values of $\sigma_k$ of corresponding periodic trajectories  
for $M=1$ and $\mathrm{S}=n/2$ where $n\in\{2,3,\dots,7\}$.}
\label{tab:1}
\end{table}
\begin{table}[t]
\centering
\begin{tabular}{c|c|c|c|c|c|c|}
\cline{2-7}
$\;$&
$\mathrm{S}=1$&
$\mathrm{S}=3/2$& $\mathrm{S}=2$&$\mathrm{S}=5/2$&$\mathrm{S}=3$&$\mathrm{S}=7/2$\\ \hline
\multicolumn{1}{|c|}{$N=3$}&-0.49&--&--&--&--&--\\ \hline
\multicolumn{1}{|c|}{$N=5$}&--&--&--&--&--&--\\ \hline
\multicolumn{1}{|c|}{$N=7$}&0.73&--&--&--&--&--\\ \hline
\multicolumn{1}{|c|}{$N=9$}&--&0.83&0.53&0.38&0.3&0.25 \\ \hline
\multicolumn{1}{|c|}{$N=11$}&--&--&0.87&0.72&0.64&0.59\\ \hline
\multicolumn{1}{|c|}{$N=13$}&--&--&--&0.9&0.82&0.77 \\ \hline
\multicolumn{1}{|c|}{$N=15$}&--&--&--&--&0.92&0.87\\ \hline
\multicolumn{1}{|c|}{$N=17$}&--&--&--&--&0.98&0.93\\ \hline
\multicolumn{1}{|c|}{$N=19$}&--&--&--&--&--&0.98\\ \hline
\end{tabular}
\caption{Values of  $\sigma_k$ corresponding periodic orbits for $M=2$ 
and $\mathrm{S}=n/2$ where $n\in\{2,3,\dots,7\}$.}
\label{tab:2}
\end{table}
Using $\sigma_k$ ($\sigma_0$) and the constraint (\ref{phi sigma const}) one obtains the equation for the mass spectrum of the system
\begin{eqnarray}
m&=-\frac{h}{g}\mathrm{S}\sigma_k+\sqrt{\left(\frac{h}{g}\right)^2 \mathrm{S}^2+m_0^2},
\label{mass sigma k}
\end{eqnarray}
or equivalently
\begin{eqnarray}
m=-\frac{h}{g}\mathrm{S}\sigma_0+\sqrt{\left(\frac{h\sigma_0}{g}\right)^2 \mathrm{S}^2+m_0^2}.
\label{mass sigma zero}
\end{eqnarray}
Since  $\sigma_k,\sigma_0\in\langle -1,1\rangle$ the mass spectrum is continuous even for fixed value of $\mathrm{S}$.
\begin{figure}[ht]
\begin{center}
\includegraphics[width=14.cm]{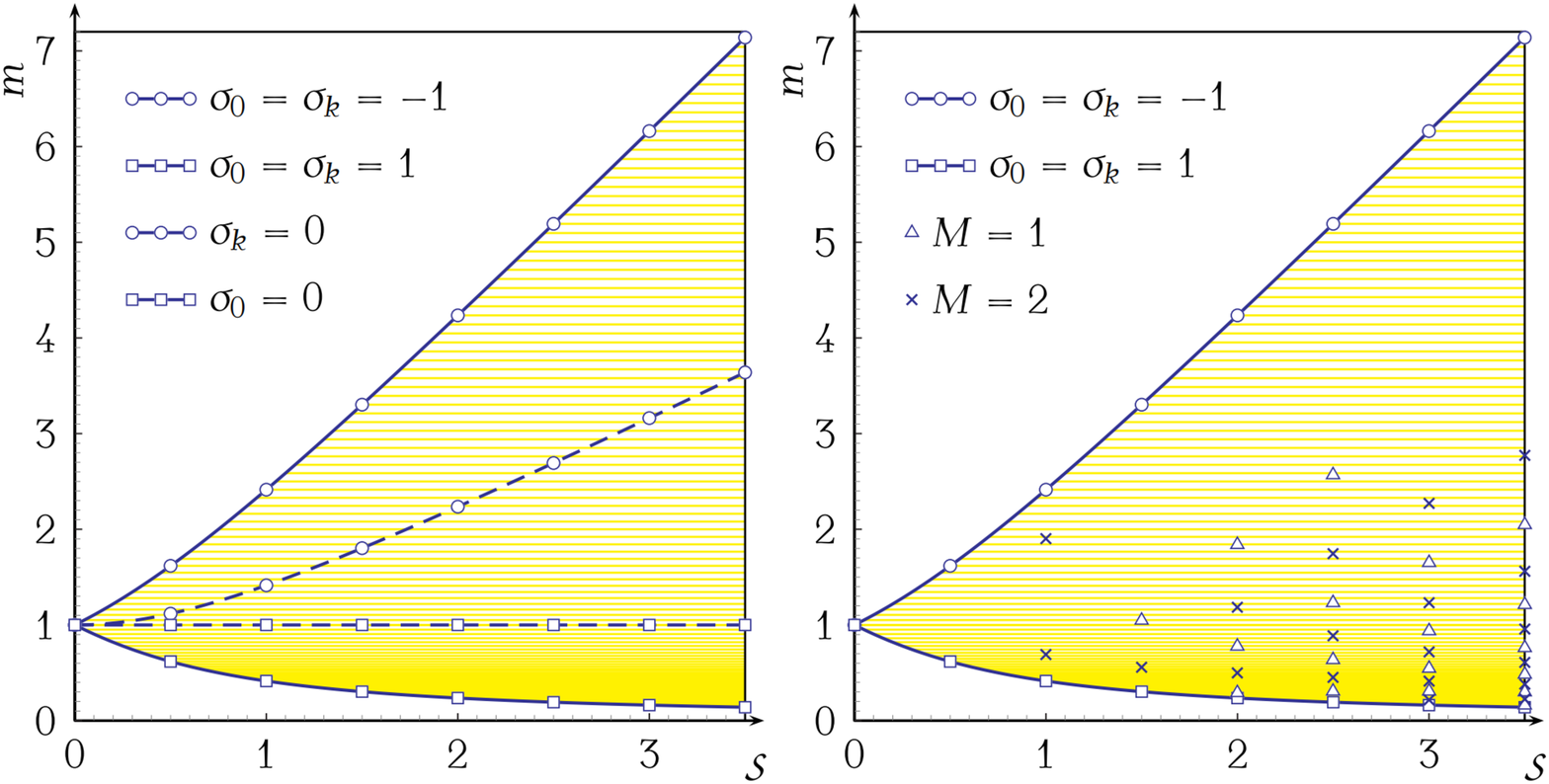}
\caption{Continuous mass spectrum for $\sigma_k\in\langle-1,1\rangle$. 
The  points marked by triangles and crosses correspond to the values 
 of $\mathrm{S}$ i $\sigma_k$ from tables  \ref{tab:1} and \ref{tab:2} accordingly. The graph was generated with  $m_0=g=h=1$.}\label{fig.2.8}
\end{center}
\end{figure}
The results can be transferred to the frame with $\vec{p}\neq 0$ by applying appropriate Lorenz boost. The analysis in such a frame does not provide any new qualitative information but complicates the picture instead.
\subsection{Luxon}
\begin{figure}[ht]
{\center
\includegraphics[width=15.cm]{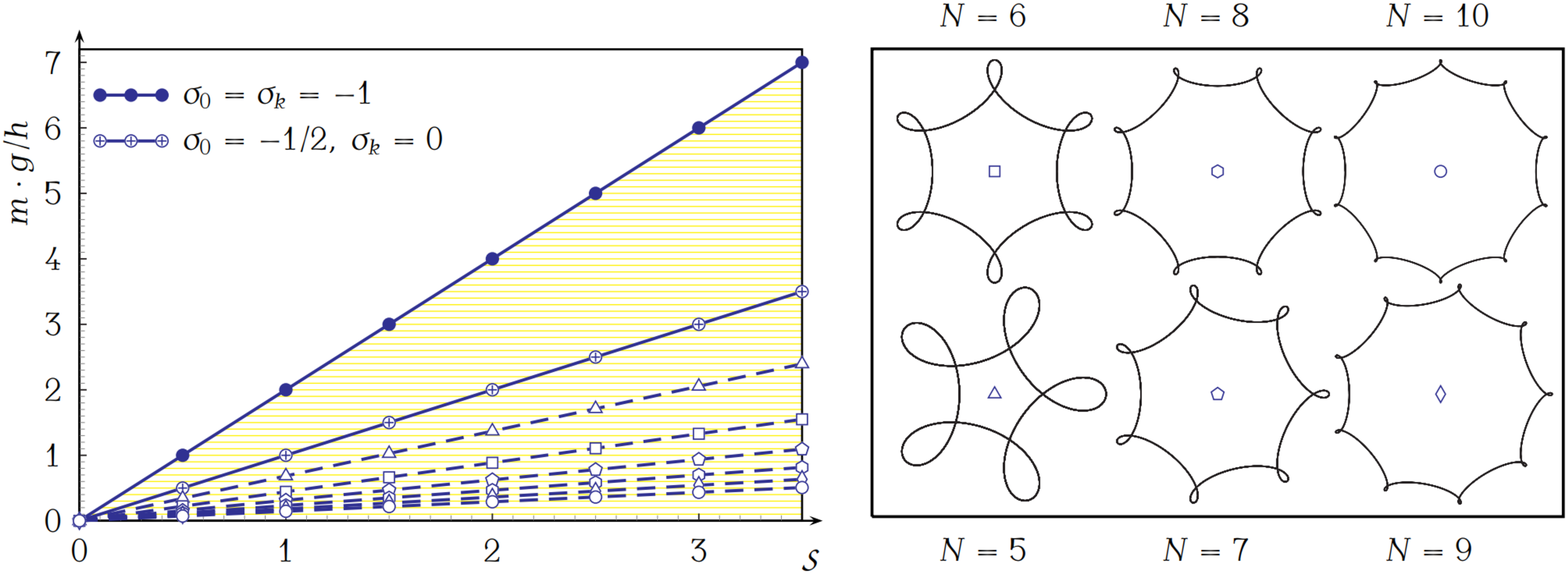}
}
\caption{Continuous mass spectrum of luxon for $\sigma_k\in\langle-1,1\rangle$. The graph on the left-hand side demonstrates first  six ($N\in\{5,6,\dots,10\}$) massive trajectories corresponding to closed orbits of type $M=1$. They are depicted on the right-hand side.}
\label{fig.2.11}
\end{figure}

Besides of the states where the spinning particle moves around mass centre with the speed lower than the speed of light 
there are motions of luxon type in the system. Assuming $m_0=0$, the four-velocity $\dot{\mathrm{x}}^\mu$ becomes light-like vector just as in the system with single spinor. The speed of such a particle is constant and equals to the speed of light:
\begin{eqnarray}
\mathrm{v}_{(l)}=|\dot{\vec{\mathrm{x}}}|/\dot{\mathrm{x}}^0=1\label{eq:lux:1}.
\end{eqnarray}
The most interesting is a luxon in configuration $\vec{\mathrm{S}}_1\cdot\vec{\mathrm{S}}_2=\mathrm{S}_1\mathrm{S}_2 $.\\
From the equation (\ref{mass sigma zero}) it follows
\begin{eqnarray}
m_l=-2\frac{h}{g}\mathrm{S}\sigma_0,\quad\mathrm{where:}\quad \sigma_0=\frac{1}{2}(\sigma_k-1)<0.
\label{masa luxon}
\end{eqnarray}
Since mass depends linearly on spin the trajectories of luxon particles are spin independent. According to (\ref{norms}):
\begin{eqnarray}
\mathrm{r}_{l}=-\frac{g}{2h\sigma_0}\sqrt{1-(1-\sigma^2)\sin^2\frac{\phi}{2}},
\end{eqnarray}
where by virtue of   (\ref{phi dot}) the parameter $\phi$ does satisfy the equation
\begin{eqnarray}
\frac{d}{d\tau_p}\phi=-2\frac{h}{g}\frac{\sigma(\phi) \sin^2\frac{\phi}{2}-2\sigma_0}{\sigma(\phi)-2\sigma_0},\label{phi :luxon}
\end{eqnarray}
and 
\begin{eqnarray}
\sigma(\phi)=\frac{2\sigma_0}{\sin^2\frac{\phi}{2}} \left[1-\sqrt{1-\left(1-\frac{\sin^2\frac{\phi}{2}}{4\sigma_0^2}\right)\sin^2\frac{\phi}{2}}\right].
\label{sigma:luxon}
\end{eqnarray}
In contrast to the case of bradyons the infinite families of closed trajectories are present here.
This conclusion follows from the fact that the quotient $N/M$ (\ref{orbity}) is growing function for $\sigma_k\in(0,1)$
with infinite at  $\sigma_k\rightarrow 1$. \\
It is worth to mention that the quotient $N/M$ does not depend on the parameters  $h$ and $g$.
Few trajectories (spin independent) of of luxons are drawn on the fig.\ref{fig.2.11}.

\section{Final remarks}
The spinning systems presented in this paper seem to have promising properties to start a discussion of the properties of the matter created in the early Universe. The asymmetric mass spectrum of the spinning particles might be a source of observed lack of balance between matter and anti-matter. Spinning particles are in fact orphaned as shown above.
It can also be a reason why the stable particles of spin higher than 1 are absent - the gap in the masses of the states of  particles and anti-particles with higher spins may be to big to create the pairs.  In effect the higher spin particles remain orphaned and are hidden elsewhere without possibility to interact with visible matter. \\
Their presence "elsewhere" can also be some explanation of the "dark energy" and "dark matter". It is also possible to suppose that "dark entities" are formed by fuzzy continuous states of spinning system.\\
In order to make the speculative remarks more serious one should not consider the spinning particles in Minkowski space but in some reasonable cosmological background instead. This is much more demanding task than the one executed here which appeared enough complicated.  
\eject

\end{document}